\documentclass[10pt, conference, letterpaper]{IEEEtran}

\usepackage{amsmath,amsfonts}
\usepackage{algorithmic}
\usepackage{algorithm}
\usepackage{array}
\usepackage[caption=false,font=normalsize,labelfont=sf,textfont=sf]{subfig}
\usepackage{textcomp}
\usepackage{stfloats}
\usepackage{url}
\usepackage{verbatim}
\usepackage{graphicx}
\usepackage{cite}
\usepackage{comment}
\usepackage{pifont}

\usepackage{multirow}
\usepackage{booktabs}
\usepackage[table,xcdraw]{xcolor}
\usepackage{tabularx}
\usepackage[hidelinks]{hyperref}
\usepackage{xurl}

\begin{document}


\title{Adaptive Traffic Camouflage: Causal and Resource-Aware Defense Against IoT Device Fingerprinting}

\author{
\IEEEauthorblockN{
Daniel Adu Worae\IEEEauthorrefmark{1},
Spyridon Mastorakis\IEEEauthorrefmark{2},
Nuno Moniz\IEEEauthorrefmark{1},
Nitesh V. Chawla\IEEEauthorrefmark{1}
}

\IEEEauthorblockA{
\IEEEauthorrefmark{1}University of Notre Dame, Notre Dame, IN, USA\\
\texttt{\{dworae,nmoniz2,nchawla\}@nd.edu}
}

\IEEEauthorblockA{
\IEEEauthorrefmark{2}Microsoft Research\\
\texttt{smastorakis@microsoft.com}
}
}

\maketitle

\begin{abstract}
Encryption hides IoT payloads, but not the traffic shape that can reveal
device identity through packet sizes, timing, direction, and
packetization. Traffic camouflage can suppress this leakage, yet a
practical defense must decide when camouflage is warranted, which
transformation to apply under bandwidth and latency constraints, and
whether its protection persists once an attacker observes defended
traffic. We present Adaptive Traffic Camouflage, a causal,
leakage-aware controller that characterizes traffic-shape leakage
without requiring runtime device labels and selects a budget-feasible
transformation for the next traffic window using confidence-gated
previous-window context. The controller chooses among heterogeneous
padding, packet-splitting, timing, and composite transformations, while
leaving traffic unchanged when camouflage is unnecessary.
We evaluate the design on three IoT device-traffic datasets,
CIC-IoT-2022, IoT Sentinel, and UNSW, using classical and
sequence-based fingerprinting models under clean-trained,
defense-aware, and incremental-exposure settings, and compare it
with fixed, random, and mean-bandwidth-matched baseline defenses.
Under the Balanced profile, camouflage reduces mean Macro-F1 by
13.2--23.3\% relative to clean traffic across the three datasets,
while incurring 4.88--7.47\% average bandwidth overhead and at most
0.64\,ms added latency. With the larger Privacy budget, the reduction
increases to 28.0--43.5\%, demonstrating substantially stronger
protection when additional communication resources are available. Defense-aware training recovers much of the lost attacker performance on CIC-IoT-2022 and UNSW, whereas IoT Sentinel retains a substantial privacy gap. A non-causal same-window
reference provides only modest additional benefit over previous-window
control, while metadata-rich attackers remain effective outside the
targeted traffic-shape surface. These results show that resource-aware
adaptive camouflage can reduce IoT traffic-shape fingerprintability
under explicit communication constraints, while the persistence of
that protection depends on how readily the defended distribution can
be learned.
\end{abstract}



\section{Introduction}
\label{sec:introduction}

Encryption protects IoT payloads, but it does not conceal the structure
of network communication. Packet sizes, timing, direction, volume, and
packetization can remain sufficiently distinctive to identify devices
and, in some settings, reveal device states or user activities
\cite{msadek2019fingerprinting,acar2020peekaboo,
wang2020fingerprinting,dong2020smarthome}. Traffic analysis therefore
creates a privacy channel even when application contents remain
inaccessible.

Traffic obfuscation addresses this leakage by modifying the observable
characteristics on which fingerprinting models rely. Existing defenses
use mechanisms such as padding, traffic shaping, packet restructuring,
timing perturbation, cover traffic, and traffic morphing
\cite{apthorpe2019smarter,perera2022iot,wang2017walkietalkie,
shenoi2023ipet,worae2025hiding}. Their effectiveness is closely coupled
to resource cost. Stronger transformations can suppress more
identifying structure, but may consume additional bandwidth or
introduce latency. Moreover, IoT leakage is multidimensional:
modifying one traffic characteristic can leave other discriminative
structure available to the attacker
\cite{engelberg2022classification,alshehri2020tunneled}. Protection
must therefore account for both what is currently leaking and what
resources can reasonably be spent to conceal it.

\footnote{\textbf{Code and results:} The implementation and evaluation artifacts are available at
\textcolor{blue}{\href{https://github.com/WadElla/Adaptive-Traffic-Camouflage}
{\nolinkurl{github.com/WadElla/Adaptive-Traffic-Camouflage}}}.}

Recent defenses have introduced dynamic or learned traffic
transformation, including learned morphing, adaptive dummy traffic,
traffic mixing, dynamic packet-length modification, and programmable
data-plane defenses
\cite{tan2024glimpse,santos2023dynamic,li2024homesentinel,
hu2025topldm,wang2025minos,nasr2021blind}. These systems demonstrate
the value of adapting traffic protection, but leave a broader control
question: \emph{given a set of heterogeneous camouflage mechanisms and
a deployment budget, which protection should be applied to the next
traffic interval, and when is protection warranted at all?} Addressing
this question also requires causal decision making. A controller cannot
observe the complete traffic characteristics of a transmitted window
and then retroactively choose how that same traffic should have been
protected.

We present Adaptive Traffic Camouflage, a causal, leakage-aware, and
resource-aware controller for IoT traffic-shape privacy. After a traffic
window completes, the controller estimates how identifying information
is distributed across packet size, timing, burst/volume, and
direction/packetization characteristics. The selected resource profile
first restricts the action space to transformations whose calibrated
bandwidth and latency costs are feasible. A global selector then matches
the observed leakage to these actions, while a context-aware model uses
richer previous-window information to predict their effectiveness on
the following window. A confidence guard permits a contextual override
only when its predicted benefit is sufficiently supported. The
controller may also leave traffic unchanged when camouflage is not
warranted. Consequently, the action for window $t$ depends only on
information available through window $t-1$.

We evaluate the design on CIC-IoT-2022, IoT Sentinel, and UNSW using
classical and sequence-based attackers under unaware, defense-aware,
fresh-realization, and incremental-exposure settings. Under the Balanced profile, camouflage reduces mean Macro-F1 by 13.2--23.3\% relative to unmodified traffic across the three datasets, while incurring 4.88--7.47\% average bandwidth overhead and at most 0.64\,ms added latency. Under the larger Privacy profile, the relative
reduction increases to 28.0--43.5\%. Defense-aware evaluation shows that the persistence of this protection is dataset-dependent: IoT Sentinel retains a meaningful privacy gap, while CIC-IoT-2022 and UNSW become substantially more learnable after the attacker observes defended traffic.

The main contributions are:

\begin{itemize}

    \item We formulate IoT traffic camouflage as a causal,
    resource-aware decision problem and develop a runtime leakage
    characterization mechanism that selects among different camouflage
    mechanisms and operating levels, including selected combinations and
    the option to leave traffic unchanged, without requiring the true
    device identity.

    \item We develop a confidence-gated causal controller that combines
    leakage-aware global selection with context-specific next-window
    gain prediction, while restricting candidate actions according to
    explicit bandwidth and latency constraints.

    \item We evaluate the controller across three public IoT datasets
    against unaware and defense-aware classical and sequence attackers,
    including incremental exposure, and compare it with fixed, random,
    and mean-bandwidth-matched defenses. The evaluation characterizes
    both the privacy achievable under different resource budgets and
    the conditions under which defended traffic remains learnable.

\end{itemize}

\section{Related Work}
\label{sec:related}

Traffic-analysis defenses reduce fingerprintability by modifying
observable communication patterns. Prior work has explored traffic
morphing, padding, packet restructuring, timing perturbation, and
related mechanisms for IoT and other encrypted traffic
\cite{wang2017walkietalkie,hafeez2019morphing,pinheiro2021adaptive,
perera2022iot,worae2025hiding}. These studies establish both the
privacy benefit of traffic transformation and its associated bandwidth
and latency costs. They also show that protection depends on which
traffic characteristics are modified, motivating defenses that can
address multiple sources of leakage rather than relying on a single
persistent transformation.

A related line of work incorporates traffic dynamics and communication
cost directly into the defense. Privacy-preserving IoT traffic shaping
has been formulated around explicit tradeoffs among privacy, dummy
traffic, and packet delay \cite{xiong2022shaping}, while smart-home
systems combine learned traffic signatures with artificial traffic
injection and partial reshaping under practical bandwidth constraints
\cite{yu2021privacyguard}. Dynamic shaping has also combined
device-specific traffic signatures with dummy-packet generation and
link padding to conceal activity-related traffic patterns at low
overhead \cite{brahma2022contextual}. More recent approaches adapt
dummy-traffic levels to network conditions \cite{santos2023dynamic},
learn traffic-morphing policies from observed flows
\cite{tan2024glimpse}, or generate configurable adversarial
perturbations under tunable communication overhead
\cite{shenoi2023ipet}. Together, these studies demonstrate that traffic
protection can benefit from adapting its behavior to traffic conditions
and resource requirements. Their adaptation, however, is primarily
defined within a particular shaping, padding, injection, or
perturbation mechanism.

Recent systems broaden the set of transformations and move traffic
protection closer to practical network deployment. Synthetic traffic
can be learned and mixed with legitimate communication
\cite{li2024homesentinel}, packet-length distributions can be modified
dynamically \cite{hu2025topldm}, and programmable data planes can
coordinate padding, dummy traffic, and scheduling
\cite{wang2025minos}. Learning-guided in-network obfuscation further
combines packet fragmentation and insertion while jointly considering
attack evasion and bandwidth overhead, including evaluation against
IoT fingerprinting attacks \cite{xie2026securitas}. These systems
provide increasingly flexible and efficient ways to construct
camouflaged traffic. Our work addresses the complementary problem of
how to allocate such protection at runtime. The controller
characterizes fingerprinting leakage in the completed traffic window,
restricts the candidate actions using explicit bandwidth and latency
constraints, and selects among heterogeneous mechanisms, operating
levels, selected compositions, or no transformation for the following
window. Thus, the adaptation concerns not only the parameters of a
camouflage mechanism, but also which resource-feasible form of
protection should be applied given the observed leakage, using only
information available through window $t-1$ to protect window $t$.

A separate concern is whether the traffic distribution created by a
defense remains protective after an attacker learns it. Prior studies
have shown that retraining and deep traffic-analysis models can
substantially weaken lightweight defenses
\cite{sirinam2018deep,mathews2023sok,nasr2021blind}, and recent IoT
defenses increasingly consider attackers with access to protected
traffic \cite{shenoi2023ipet,hu2025topldm}. This concern is especially
important for adaptive camouflage because changing the transformation
policy does not by itself prevent the resulting defended distribution
from becoming learnable. We therefore evaluate more than the immediate
degradation of a clean-trained classifier. Each defense is tested
against an attacker trained on its corresponding defended traffic,
followed by fresh-realization and incremental-exposure experiments.
Fixed, randomized, and mean-bandwidth-matched controls separate
adaptive allocation from action diversity and resource expenditure,
while metadata-rich attackers expose identifying signals outside the
traffic-shape surface targeted by the controller.

\section{Threat Model and Problem Formulation}
\label{sec:threat_model}

\subsection{Adversary Model}

We consider a passive adversary whose objective is to identify an IoT
device from encrypted network traffic. The adversary observes traffic
downstream of the camouflage layer and may record and analyze packet
traces, but does not inject, modify, delay, or drop packets, nor
compromise communicating endpoints. Application payloads are assumed
to be encrypted and unavailable to the adversary. We consider a
closed-set device-identification setting in which the target belongs to
the set of device classes represented during attacker training.

We distinguish two adversarial observation surfaces. The
\emph{traffic-shape surface}, which defines the primary protection
objective of this work, comprises packet and wire sizes, timing and
inter-arrival behavior, traffic direction, packetization and burst
structure, traffic volume, and the resulting ordered packet sequence.
These characteristics remain observable despite payload encryption and
can provide discriminative signatures for device identification.

We further consider an \emph{extended metadata surface} that augments
traffic shape with network-, transport-, and flow-level information
available at the observation point, including protocol composition,
port and service-related information, endpoint characteristics, packet
flags, and flow statistics. The camouflage mechanism does not
systematically rewrite all such semantics; we use this observation
surface to quantify residual leakage beyond the traffic characteristics
directly targeted by the defense.

We do not assume that the defense mechanism is secret. The adversary
may know the camouflage strategy, the available transformations, and
the defender's resource constraints, and may obtain defended traffic
to train models on the distribution induced by the defense.

\subsection{Defender Model}

The defender is a trusted traffic-processing component that applies
camouflage to protected IoT traffic before it reaches the adversary's
observation point. The defense may be realized at the IoT endpoint or
within a trusted network component along the communication path,
depending on the deployment environment.

The defender does not require application payload contents or the true
device identity when making runtime decisions. Instead, it
characterizes recently observed traffic and estimates the extent and
type of traffic-shape leakage before deciding whether camouflage is
warranted and which resource-feasible protection should be applied.
The defender may also leave a traffic window unchanged when protection
is not warranted.

A central requirement is \emph{causal operation}. Let $W_t$ denote the
traffic transmitted during window $t$, and let $\mathbf{z}_{t-1}$
represent the traffic-shape and leakage characterization derived from
the preceding window. For a window with valid predecessor context, the
action applied to $W_t$ is selected as

\begin{equation}
    a_t = \pi(\mathbf{z}_{t-1}, \mathbf{B}),
    \qquad a_t \in \mathcal{A}_{\mathbf{B}},
    \label{eq:causal_policy}
\end{equation}

where $\pi$ denotes the camouflage policy, $\mathbf{B}$ represents the
operator-specified communication constraints, and
$\mathcal{A}_{\mathbf{B}}$ denotes the set of actions deemed feasible
under those constraints. If no valid predecessor is available, the
controller falls back to a calibration-derived global decision rather
than using information from the current window. Thus, the action for
$W_t$ is always determined from information available before that
window is transmitted.

\subsection{Privacy Objective and Scope}

The defender seeks to reduce device-identifying information on the
traffic-shape surface while respecting explicit communication
constraints. Because both leakage and the privacy--cost characteristics
of available actions can vary across traffic windows, the objective is
to select a resource-feasible action when protection is warranted,
using only causally available observations.

We evaluate privacy through the performance of device-identification
adversaries on held-out traffic, using Macro-F1 as the primary metric;
lower Macro-F1 indicates greater suppression of device-identifying
information. Our objective is empirical rather than
information-theoretic: we do not claim anonymity or a formal privacy
guarantee. Endpoint compromise, payload disclosure, active traffic
manipulation, and complete concealment of protocol, service, and flow
semantics are outside the protection objective. Instead, we study
whether causal, resource-constrained traffic camouflage can reduce IoT
device fingerprintability on the traffic-shape surface, including when
the adversary is permitted to learn from defended traffic.

\section{Adaptive Camouflage Design}
\label{sec:design}

\subsection{Design Overview}
\label{sec:design_overview}

\begin{figure*}[!t]
    \centering  \includegraphics[width=1.0\textwidth]{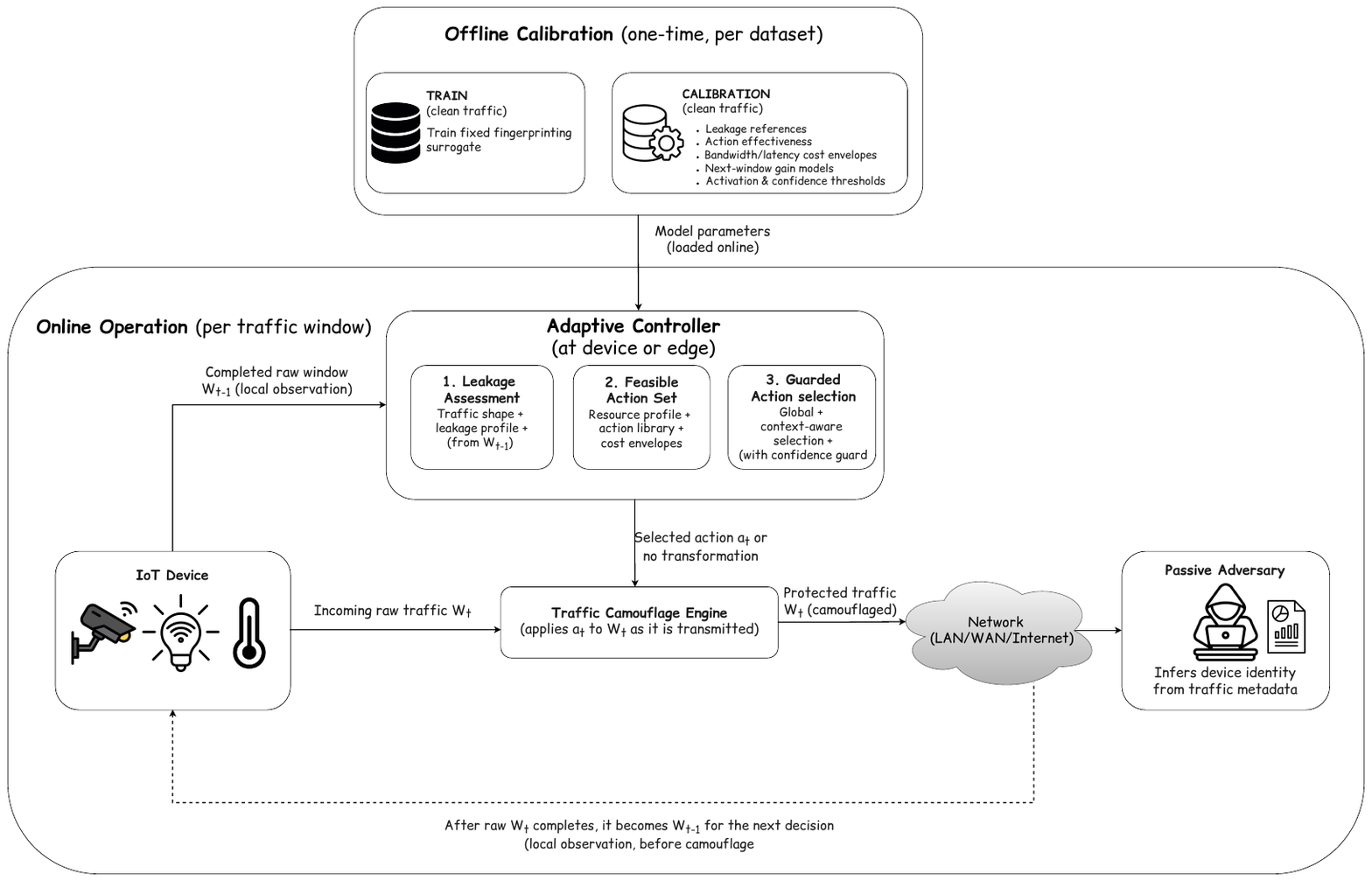}
    \caption{Architecture of Adaptive Traffic Camouflage. Offline calibration
    derives leakage references, action-effectiveness and resource-cost
    estimates, next-window gain models, and decision thresholds. During
    online operation, the completed window $W_{t-1}$ is used to select a
    resource-feasible action for $W_t$, which is camouflaged as it is
    transmitted before reaching the adversary.}
    \label{fig:architecture}
\end{figure*}

Adaptive Traffic Camouflage operates over fixed, non-overlapping 10-s
traffic windows and follows the causal policy defined in
Eq.~\ref{eq:causal_policy}. As illustrated in
Fig.~\ref{fig:architecture}, the design separates offline calibration
from online traffic protection.

Offline calibration establishes the information required for runtime
decision-making. A fingerprinting surrogate is trained on clean
\textsc{Train} traffic, while the disjoint \textsc{Calibration}
partition is used to characterize leakage patterns, estimate the
privacy effectiveness and communication cost of the available
camouflage actions, train the next-window gain models, and determine
the controller's activation and confidence thresholds. These
calibration-derived quantities are established before evaluation and
are not updated using \textsc{Test} traffic.

During online operation, the controller analyzes the completed
preceding window to estimate its traffic-shape leakage. The selected
resource profile restricts the available actions to those that satisfy
the corresponding bandwidth and latency constraints. From this
feasible set, the controller combines leakage-aware global selection
with context-aware prediction of next-window effectiveness, while a confidence guard determines whether the contextual recommendation
should be accepted. The resulting action, including the
possibility of leaving traffic unchanged, is applied to the following
window as its packets arrive.

\subsection{Runtime Leakage Characterization}
\label{sec:leakage}

Each completed traffic window is summarized by 24 traffic-shape
features organized into four semantic groups: packet size, timing,
burst and volume, and direction and packetization. These features
describe packet-length distributions, inter-arrival behavior, traffic
intensity, directional balance, and packetization changes. Protocol,
port, service, endpoint, and other richer network semantics are not
used by the controller.

The leakage profiler uses a fingerprinting surrogate as a measurement
instrument. The surrogate is trained once on clean \textsc{Train}
traffic using the same 24-dimensional traffic-shape representation
and is then held fixed. Although device labels are required to train
this surrogate offline, neither the profiler nor the deployed
controller receives the true device identity at runtime. The profiler
therefore measures which parts of the observed traffic support the
surrogate's current fingerprinting hypothesis rather than requiring
the true device label of the window being analyzed.

Let $\mathbf{x}$ denote the traffic-shape representation of a completed
window, and let

\begin{equation}
    c^* = \arg\max_c p(c\mid\mathbf{x})
    \label{eq:surrogate_class}
\end{equation}

denote the class receiving the highest support from the fixed
surrogate. The profiler does not treat $c^*$ as ground truth. Instead,
it asks which semantic groups of traffic features are responsible for
supporting that particular hypothesis.

For each semantic group $g$, the profiler replaces only that group's
features with reference blocks drawn from real
\textsc{Calibration} traffic while leaving all remaining groups
unchanged. It then measures the reduction in the surrogate's support
for the same original class $c^*$:

\begin{equation}
    L_g(\mathbf{x}) =
    \left[
    p(c^*\mid\mathbf{x}) -
    \mathbb{E}_{\mathbf{r}\sim\mathcal{R}}
    \left[
    p\!\left(
        c^* \mid
        \mathbf{x}_{g\leftarrow\mathbf{r}_g}
    \right)
    \right]
    \right]_+ ,
    \label{eq:leakage}
\end{equation}

where $\mathcal{R}$ is a weighted empirical reference distribution
constructed from real \textsc{Calibration} traffic, $\mathbf{r}$ is a
reference traffic-shape vector drawn from this distribution, and
$\mathbf{r}_g$ denotes its feature block for group $g$. The notation
$\mathbf{x}_{g\leftarrow\mathbf{r}_g}$ denotes the representation
obtained by replacing only group $g$ in $\mathbf{x}$ with the
corresponding reference block, while $[\cdot]_+$ retains only positive
reductions in surrogate support. Because the replacement blocks are
taken from observed calibration traffic, the intervention preserves
the joint structure of related features within each group.

A large $L_g(\mathbf{x})$ therefore means that replacing the
window's original characteristics in group $g$ substantially weakens
the surrogate's current hypothesis. The four values form the leakage
profile
$\mathbf{L}(\mathbf{x})=\{L_g(\mathbf{x})\}_{g=1}^{4}$, while
$L(\mathbf{x})=\sum_g L_g(\mathbf{x})$ denotes aggregate leakage.
The vector identifies where the fingerprinting signal is concentrated;
the aggregate score determines whether the observed leakage is large
enough to justify camouflage.

\subsection{Resource-Constrained Action Space}

The controller selects from a heterogeneous action space spanning
packet-size shaping, packet splitting and packetization, timing
modification, cover traffic, multiple operating strengths, and selected
cross-dimension compositions. A no-transformation option is retained as
an explicit action, allowing the controller to avoid spending resources
when leakage is low or no feasible transformation offers a meaningful
predicted improvement. The action space therefore captures both how
traffic should be modified and how strongly the modification should be
applied.

The controller is instantiated with 17 camouflage configurations and an
explicit no-transformation action. The library includes padding at 5\%,
10\%, 20\%, and 35\%; packet splitting at 5\%, 10\%, and 20\%;
micro-batching at 10, 25, and 50\,ms; cover traffic at 5\%, 10\%, and
20\%; and four cross-dimension compositions combining 5\% padding,
splitting, or cover traffic with 10-ms micro-batching, or 5\% padding
with 5\% splitting. These configurations define the candidate action
library, while the actions available under a particular operating
profile are determined separately using the calibration-derived
bandwidth and latency constraints in Eq.~\eqref{eq:feasible_actions}. An
action may therefore remain part of the library while being excluded
from a given profile when its measured cost exceeds that profile's
resource limits.

Resource constraints are enforced before action selection. We consider
three deployment profiles: Tight, which permits up to 5\% bandwidth
overhead and 10\,ms additional latency; Balanced, which permits 10\%
and 25\,ms, respectively; and Privacy, which permits 20\% and 50\,ms.
Balanced serves as the primary operating point, while the other profiles
expose the privacy--cost tradeoff.

For each candidate action, communication costs are estimated from
\textsc{Calibration} traffic. For each device, we compute the
95th-percentile bandwidth overhead and added latency produced by the
action, and use the largest device-level value for each resource as a
conservative cost estimate. Given a resource profile $\mathbf{B}$, the
feasible action set is

\begin{equation}
    \mathcal{A}_{\mathbf{B}}
    =
    \left\{
        a \in \mathcal{A} :
        C^{\mathrm{bw}}_a \leq B^{\mathrm{bw}},
        \;
        C^{\mathrm{lat}}_a \leq B^{\mathrm{lat}}
    \right\},
    \label{eq:feasible_actions}
\end{equation}

where $C^{\mathrm{bw}}_a$ and $C^{\mathrm{lat}}_a$ denote the
conservative calibration-derived bandwidth and latency costs of action
$a$. Resource feasibility is evaluated before privacy-based action
selection, so an action that exceeds either constraint is excluded
rather than traded against privacy benefit through a weighted
objective.

\subsection{Global Selection and Context-Aware Reranking}
\label{sec:selection}

The global selector considers only actions that satisfy the selected
resource profile. For each action $a$ and leakage group $g$,
\textsc{Calibration} traffic is used to estimate the fractional
reduction in group leakage produced by the action. We use a conservative
lower-bound estimate of this effect, denoted by $E_{a,g}$, so that
actions whose benefit is uncertain or that amplify leakage are not
favored by an optimistic average estimate. Let
$\mathbf{L}_{t-1}$ denote the four-dimensional leakage profile of the
preceding window. For each feasible action $a$, the predicted residual
leakage is

\begin{equation}
    R(a\mid\mathbf{L}_{t-1})
    =
    \sum_g
    L_{t-1,g}\left(1-E_{a,g}\right),
    \label{eq:residual}
\end{equation}

where $L_{t-1,g}$ is the leakage attributed to group $g$ in
$W_{t-1}$. Thus, actions that are effective against the leakage
components that dominate the preceding window receive smaller predicted
residual-leakage values.

When the aggregate leakage exceeds the calibration-derived activation
threshold, the global candidate is

\begin{equation}
    a_t^{G}
    =
    \arg\min_{a\in\mathcal{A}_{\mathbf{B}}}
    R(a\mid\mathbf{L}_{t-1}),
    \label{eq:global_choice}
\end{equation}

where $\mathcal{A}_{\mathbf{B}}$ is the resource-feasible action set
defined in Eq.~\ref{eq:feasible_actions}. If aggregate leakage does not
exceed the activation threshold, or if the best feasible camouflage
action provides no positive predicted reduction, the controller selects
no transformation. When multiple actions provide effectively the same
predicted privacy benefit, preference is given to the action that places
less pressure on the available resources.

The context-aware reranker addresses a different question. Two windows
can exhibit similar aggregate leakage profiles while differing in their
detailed packet-size, timing, volume, and directional characteristics.
The reranker therefore uses the complete traffic-shape context of the
preceding window rather than only its group-level leakage profile. We
define this context as

\begin{equation}
    \mathbf{z}_{t-1}
    =
    \left[
        \mathbf{x}_{t-1},
        \mathbf{L}_{t-1},
        L_{t-1}
    \right],
    \label{eq:context_vector}
\end{equation}

where $\mathbf{x}_{t-1}$ contains the 24 traffic-shape features,
$\mathbf{L}_{t-1}$ contains the four group-leakage values, and
$L_{t-1}$ is aggregate leakage. Thus,
$\mathbf{z}_{t-1}$ is a 29-dimensional representation of the
preceding traffic context and contains neither the true device identity
nor information from the upcoming window.

The gain models are trained exclusively from valid consecutive-window
pairs in \textsc{Calibration}. Consider a pair
$W_{j-1}\rightarrow W_j$. The input is the preceding-window context
$\mathbf{z}_{j-1}$. To construct the target for a candidate action
$a$, the action is applied to the following window $W_j$. Let
$\mathbf{x}_j$ denote the original traffic-shape representation of
$W_j$, let $c_j^*$ denote the surrogate hypothesis defined in
Eq.~\ref{eq:surrogate_class}, and let
$\mathbf{x}_j^{(a,\omega)}$ denote the representation obtained after
applying action $a$ under realization $\omega$. The realized gain is

\begin{equation}
    G_{j,a}
    =
    p(c_j^*\mid\mathbf{x}_j)
    -
    \mathbb{E}_{\omega}
    \left[
        p\!\left(
            c_j^*\mid\mathbf{x}_j^{(a,\omega)}
        \right)
    \right],
    \label{eq:realized_gain}
\end{equation}

where the expectation represents the empirical average over the
calibration realizations of action $a$. A positive $G_{j,a}$ indicates
that the action reduces the surrogate's support for its original
fingerprinting hypothesis, while a negative value indicates that the
action strengthens that support.

For each resource profile $\mathbf{B}$, a separate regression model is
trained for every camouflage action that is feasible under that
profile. The model predicts an action's next-window gain from the
preceding-window context. At runtime,

\begin{equation}
    \widehat{G}_{t,a}^{(\mathbf{B})}
    =
    f_{a,\mathbf{B}}(\mathbf{z}_{t-1}),
    \label{eq:context_gain}
\end{equation}

where $f_{a,\mathbf{B}}$ is the gain model for action $a$ under resource
profile $\mathbf{B}$. The following window is used only to construct
the supervised target during calibration; runtime predictions depend
only on information from the completed preceding window.

Let $\mathcal{A}_{\mathbf{B}}^{+}\subseteq
\mathcal{A}_{\mathbf{B}}$ denote the feasible camouflage actions
excluding the no-transformation option. When aggregate leakage exceeds
the calibration-derived activation threshold, the reranker predicts
the next-window gain of each action in
$\mathcal{A}_{\mathbf{B}}^{+}$ and proposes

\begin{equation}
    a_t^{C}
    =
    \arg\max_{a\in\mathcal{A}_{\mathbf{B}}^{+}}
    \widehat{G}_{t,a}^{(\mathbf{B})},
    \label{eq:context_choice}
\end{equation}

provided that at least one feasible camouflage action has positive
predicted gain. If the activation condition is not satisfied or no
such action exists, no contextual candidate is proposed and the global
candidate is retained. Resource feasibility remains external to the
gain predictor; an action cannot become eligible solely because it
receives a high predicted gain.

The two selectors are therefore complementary rather than redundant.
The global selector combines the current leakage composition with
action effectiveness learned across the calibration distribution to
identify the generally strongest feasible response. The context-aware
reranker instead learns from consecutive calibration windows whether
the detailed state of $W_{t-1}$ predicts that a particular feasible
action will be especially effective on the following window. The confidence guard described next determines whether this contextual recommendation is sufficiently well supported to be accepted.

\subsection{Confidence-Gated Causal Execution}
\label{sec:guard}

The context-aware reranker provides a traffic-specific prediction,
but such predictions need not be equally reliable for every observed
context. The controller therefore retains the global candidate as its
default decision and accepts the contextual recommendation only when
its predicted gain is sufficiently large according to a threshold
selected from \textsc{Calibration} data.

Let $a_t^{G}$ denote the global candidate and let $a_t^{C}$ denote the
contextual candidate when one exists. The final action is

\begin{equation}
    a_t =
    \begin{cases}
        a_t^{C}, &
        a_t^{C}\ \text{exists and}\ 
        \widehat{G}_{t,a_t^{C}}^{(\mathbf{B})}
        \geq \tau_{\mathbf{B}},\\[1.5mm]
        a_t^{G}, &
        \text{otherwise},
    \end{cases}
    \label{eq:guard}
\end{equation}

where $\tau_{\mathbf{B}}$ is the confidence threshold for resource
profile $\mathbf{B}$, selected using \textsc{Calibration} data only.
The predicted gain of the contextual candidate serves directly as the
acceptance signal. If no contextual candidate is available or its
predicted gain does not reach $\tau_{\mathbf{B}}$, the global candidate
is retained.

When no valid preceding-window context is available, such as at the
beginning of a traffic stream or after a gap that invalidates the
predecessor relationship, the controller also uses the global selector
directly. Thus, context-specific adaptation is permitted only when both
a valid preceding context and sufficiently strong predicted benefit are
available.

Once $a_t$ is selected, the corresponding camouflage transformation
begins operating on $W_t$ as its packets arrive. Neither the global
selector nor the context-aware reranker observes the completed
traffic-shape statistics of $W_t$ before making this decision.
Consequently, all information used to select $a_t$ is available no
later than the completion of $W_{t-1}$, preserving the causal policy
defined in Eq.~\ref{eq:causal_policy}.

\section{Evaluation Methodology}
\label{sec:evaluation}

\subsection{Datasets and Experimental Splits}
\label{sec:datasets}

We evaluate the proposed defense on three public IoT traffic
datasets: CIC-IoT-2022~\cite{dadkhah2022ciciot}, IoT
Sentinel~\cite{miettinen2017iotsentinel}, and the UNSW IoT
dataset~\cite{sivanathan2019classifying}. Each evaluation includes six
device classes. To separate defense construction, attacker adaptation,
and final assessment, each dataset is divided into disjoint
\textsc{Train}, \textsc{Calibration}, \textsc{Adapt}, and
\textsc{Test} partitions, as summarized in
Table~\ref{tab:dataset_splits}. CIC-IoT-2022 and UNSW are partitioned
by capture date, while IoT Sentinel is partitioned by device-capture
group.

\begin{table}[t]
\centering
\caption{Dataset partitions.}
\label{tab:dataset_splits}
\footnotesize
\setlength{\tabcolsep}{3.5pt}
\renewcommand{\arraystretch}{1.08}

\begin{tabular}{@{}lcccc@{}}
\toprule
&
\multicolumn{4}{c}{\textbf{Traffic windows}} \\
\cmidrule(l){2-5}
\textbf{Dataset} &
\textbf{Train} &
\textbf{Calibration} &
\textbf{Adapt} &
\textbf{Test} \\
\midrule

CIC-IoT-2022
& 12,000 (8)
& 6,000 (4)
& 4,500 (3)
& 7,500 (5) \\

IoT Sentinel
& 507 (54)
& 216 (24)
& 178 (18)
& 232 (24) \\

UNSW
& 13,500 (9)
& 6,000 (4)
& 4,500 (3)
& 5,676 (4) \\

\bottomrule
\end{tabular}

\vspace{2pt}

\parbox{0.98\columnwidth}{\scriptsize
\textit{Note:} Values report the number of traffic windows; parentheses
report the number of independent capture units used in each partition.
}
\end{table}

\textsc{Train} is used to fit the clean fingerprinting attackers and
the fixed surrogate used by the leakage profiler.
\textsc{Calibration} is used to derive leakage references, action
effectiveness and resource costs, context-specific gain models,
decision thresholds, feasible action sets, and baseline parameters.
\textsc{Adapt} is reserved for attacker exposure to defended traffic,
while \textsc{Test} is held out for final evaluation. The controller
architecture is shared across datasets, but calibration-derived
quantities are learned independently for each dataset.

\subsection{Adversary Models and Observation Surfaces}
\label{sec:attackers}

Table~\ref{tab:attacker_regimes} summarizes the five attacker regimes.
A0 represents an attacker trained only on clean traffic. A1 evaluates a
fixed defense after the attacker has learned from ADAPT traffic protected
with that same fixed configuration. A2 is the primary defense-aware
setting. For each evaluated defense, a fresh attacker is trained from
scratch on clean TRAIN together with ADAPT traffic generated by that
same defense, and is then evaluated on TEST traffic protected by the
corresponding defense. Thus, the attacker learns the same defense
configuration or policy that it later encounters during evaluation.

A3 reuses the adaptive A2 attacker on a fresh stochastic realization of
the same adaptive policy without further training. A4 instead starts
from a clean-trained neural attacker and updates it sequentially as
defended ADAPT traffic becomes available.

\begin{table*}[!t]
\centering
\caption{Attacker regimes used in the evaluation.}
\label{tab:attacker_regimes}
\footnotesize
\setlength{\tabcolsep}{5pt}
\renewcommand{\arraystretch}{1.08}
\begin{tabular}{@{}p{0.07\textwidth}
                p{0.28\textwidth}
                p{0.27\textwidth}
                p{0.29\textwidth}@{}}
\hline
\textbf{Regime} &
\textbf{Training or Update Support} &
\textbf{Evaluation} &
\textbf{Interpretation} \\
\hline

A0 &
Clean \textsc{Train} only &
Clean and defended \textsc{Test} &
Attacker has no prior exposure to the defense. \\

A1 &
Clean \textsc{Train} and fixed-defended \textsc{Adapt} &
The corresponding fixed defense on \textsc{Test} &
Measures the learnability of a predictable fixed transformation. \\

A2 &
Clean \textsc{Train} and corresponding defended \textsc{Adapt} &
The corresponding defended \textsc{Test} &
Fresh defense-aware attacker trained separately for each evaluated
defense. \\

A3 &
Adaptive A2 attacker without additional training &
A fresh stochastic realization of adaptive \textsc{Test} &
Tests whether a new realization remains protective after the defended
distribution has been learned. \\

A4 &
Clean-trained neural attacker sequentially updated using defended
\textsc{Adapt} &
Fixed defended \textsc{Test} after successive exposure stages &
Measures attacker recovery as exposure to defended traffic increases. \\
\hline
\end{tabular}
\end{table*}

The primary traffic-shape attackers are Random Forest, Histogram
Gradient Boosting, and a multilayer perceptron operating on the
24-dimensional representation used by the controller, together with
CNN1D and Transformer models operating on ordered packet-size, timing,
and direction sequences. We additionally evaluate metadata-rich
attackers that include protocol-, endpoint-, port/service-, packet-flag-,
and flow-level information to measure leakage beyond the traffic-shape
surface targeted by the defense.

\subsection{Defense Baselines}
\label{sec:baselines}

We compare the adaptive controller with three baselines that separate
the value of leakage-aware allocation from the effects of using one
strong transformation, varying actions over time, or spending a similar
communication budget.

\paragraph{Best feasible fixed defense}
This baseline applies a single resource-feasible camouflage
configuration throughout the evaluation. Among the feasible
configurations, the one yielding the lowest surrogate Macro-F1 on
\textsc{Calibration} is selected and then held fixed for
\textsc{Adapt} and \textsc{Test}.

\paragraph{Random heterogeneous defense}
This baseline samples uniformly from the resource-feasible camouflage
actions, excluding the no-transformation option whenever at least one
camouflage action is feasible. Selection is independent of the leakage
profile and traffic context. The baseline therefore preserves action
diversity while removing leakage-aware selection, allowing us to test
whether adaptation provides benefit beyond simply varying feasible
transformations.

\paragraph{Mean-bandwidth-matched static defense}
This baseline applies a single camouflage transformation according to a
traffic-independent duty schedule selected using \textsc{Calibration}.
The schedule is chosen so that its mean bandwidth overhead does not
exceed that of the adaptive controller on \textsc{Calibration}, while
also remaining within the adaptive controller's mean latency
expenditure. The transformation and duty schedule are fixed before
\textsc{Adapt} and \textsc{Test}. This comparison separates adaptive
allocation from the effect of spending a similar average bandwidth
budget.

In the defense-aware setting, each defense is evaluated against a
separately trained attacker using clean \textsc{Train} and
\textsc{Adapt} traffic generated by that same defense. Thus, each
attacker is exposed to the defense configuration or policy it later
encounters during evaluation.

\subsection{Evaluation Metrics and Statistical Analysis}
\label{sec:metrics}

Macro-F1 is the primary device-identification metric, with lower values
indicating stronger privacy. Communication cost is measured by relative
bandwidth overhead and added latency, together with realized
resource-tier violations. We characterize controller behavior through
its action distribution, no-transformation rate, and frequency with
which the contextual recommendation passes the confidence guard.

Statistical inference respects the independent capture structure of
each dataset rather than treating individual traffic windows as
independent observations. CIC-IoT-2022 and UNSW contain five and four
independent \textsc{Test} dates, respectively, while IoT Sentinel
contains 24 device-capture groups. Paired Macro-F1 differences are
summarized using 95\% cluster-bootstrap confidence intervals over these
units and paired randomization tests, using exact enumeration for
CIC-IoT-2022 and UNSW and Monte Carlo randomization for IoT Sentinel.
Because the smallest attainable two-sided $p$-values are 0.0625 for
CIC-IoT-2022 and 0.125 for UNSW, we emphasize effect sizes, confidence
intervals, and consistency across independent capture units.

\section{Evaluation Results}
\label{sec:results}

\subsection{Privacy--Cost Tradeoff Under Unaware Attackers}
\label{sec:privacy_cost}

\noindent\textbf{RQ1: How effectively does causal adaptive camouflage
reduce traffic-shape fingerprintability under explicit resource
constraints?}

\begin{figure}[t]
    \centering
    \includegraphics[width=\columnwidth]
    {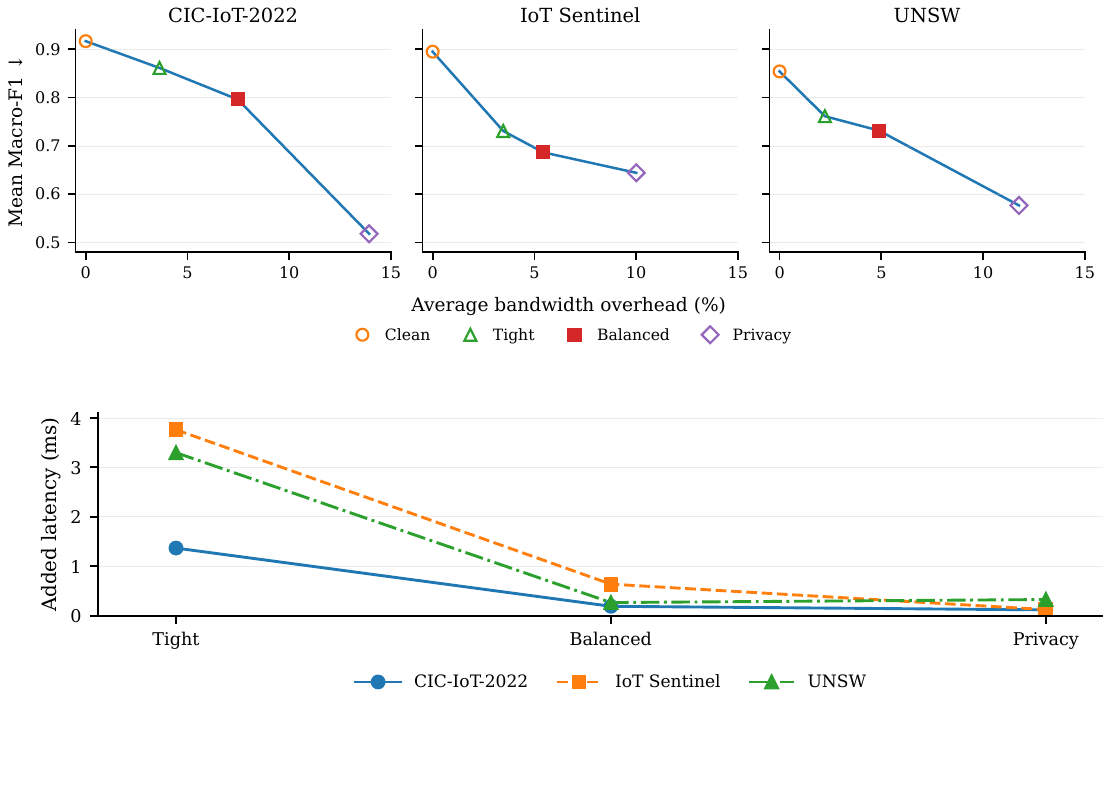}
    \caption{Privacy--resource tradeoff under an attacker trained only
    on clean traffic (A0). (a) Mean Macro-F1 across the five core
    traffic-shape attackers versus average bandwidth overhead; lower
    Macro-F1 indicates lower fingerprintability. Clean denotes
    unmodified traffic, while Tight, Balanced, and Privacy denote the
    three adaptive resource profiles. (b) Average added latency under
    the adaptive profiles. Tight minimizes bandwidth consumption but
    incurs substantially greater latency than Balanced across all three
    datasets, illustrating that the controller trades between bandwidth
    and latency rather than scaling a single resource dimension.}
    \label{fig:privacy_cost}
\end{figure}

Figure~\ref{fig:privacy_cost} summarizes the privacy--resource tradeoff
under A0, where the attacker is trained only on clean traffic. Without
camouflage, mean Macro-F1 across the five core traffic-shape attackers
is 0.917 on CIC-IoT-2022, 0.895 on IoT Sentinel, and 0.854 on UNSW.
Under the Balanced profile, these values decrease to 0.796, 0.687, and
0.732, corresponding to relative reductions of 13.2\%, 23.3\%, and
14.3\%, respectively. The associated average bandwidth overhead is
7.47\%, 5.44\%, and 4.88\%, respectively, with only
0.19--0.64\,ms of added latency.

The Tight and Privacy profiles further expose the privacy--cost
tradeoff. Tight limits average bandwidth overhead to 2.23--3.63\% but
yields smaller privacy gains and higher latency, reflecting greater use
of timing-oriented actions when bandwidth is more constrained. In
contrast, the larger Privacy budget enables stronger protection,
lowering mean Macro-F1 to 0.518, 0.644, and 0.577 on CIC-IoT-2022,
IoT Sentinel, and UNSW, respectively, at 10.01--13.93\% bandwidth
overhead. These results show that the resource profile changes which
actions and operating strengths are feasible, allowing the controller
to trade bandwidth and latency for privacy rather than simply scaling
a single defense.

The calibration-derived feasibility estimates also transfer well to
held-out traffic. Under Balanced, per-window resource-limit violations
are 0.013\% on CIC-IoT-2022, 0\% on IoT Sentinel, and 0.088\% on
UNSW; these are small bandwidth overshoots, with no latency violations.
No violations occur under Tight or Privacy. The protection level,
however, varies across attacker architectures. In particular, the
Transformer remains comparatively strong on CIC-IoT-2022 and UNSW,
where Balanced changes Macro-F1 only from 0.919 to 0.906 and from
0.804 to 0.799, respectively, despite using the same packet-shape
information as the CNN1D attacker.

\subsection{Adaptive Selection versus Fair Baseline Defenses}
\label{sec:fair_baselines}

\noindent\textbf{RQ2: Does leakage-aware adaptive selection provide a
privacy advantage over random, fixed, and mean-bandwidth-matched
baseline defenses?}

Table~\ref{tab:fair_baselines} compares the Balanced controller with
the three baselines defined in Section~\ref{sec:baselines}. The results
show that the value of adaptive selection depends on both the traffic
distribution and the effectiveness of the available actions.

\begin{table}[t]
\centering
\caption{Macro-F1 across baseline defenses.}
\label{tab:fair_baselines}
\footnotesize
\setlength{\tabcolsep}{3.4pt}
\renewcommand{\arraystretch}{1.08}

\begin{tabular}{@{}llcccc@{}}
\toprule
& &
\multicolumn{4}{c}{\textbf{Mean Macro-F1} $\downarrow$} \\
\cmidrule(l){3-6}
\textbf{Dataset} &
\textbf{Regime} &
\textbf{Adaptive} &
\textbf{Random} &
\textbf{Static} &
\textbf{Fixed} \\
\midrule

\multirow{2}{*}{CIC-IoT-2022}
& A0 & 0.796 & 0.880 & 0.851 & 0.844 \\
& A2 & 0.925 & 0.916 & 0.928 & 0.923 \\

\addlinespace[2pt]

\multirow{2}{*}{IoT Sentinel}
& A0 & 0.687 & 0.790 & 0.665 & 0.665 \\
& A2 & 0.791 & 0.833 & 0.809 & 0.823 \\

\addlinespace[2pt]

\multirow{2}{*}{UNSW}
& A0 & 0.723 & 0.781 & 0.643 & 0.706 \\
& A2 & 0.834 & 0.796 & 0.799 & 0.792 \\

\bottomrule
\end{tabular}

\vspace{2pt}

\parbox{0.98\columnwidth}{\scriptsize
\textit{Note:} Lower Macro-F1 indicates stronger privacy.
A0 uses attackers trained only on clean traffic; A2 uses fresh
defense-aware attackers trained separately for each defense.
Static uses a traffic-independent single-action schedule calibrated
to approximate the adaptive controller's mean bandwidth expenditure.
Fixed continuously applies the best feasible single action.
}
\end{table}

On CIC-IoT-2022, adaptive selection achieves the lowest A0 Macro-F1
(0.796), but the defenses converge under A2 to 0.916--0.928. The
Balanced policy is also strongly concentrated on a single padding
configuration, which may make the resulting defended distribution
easier to learn once representative protected traffic becomes
available.

IoT Sentinel provides the clearest defense-aware benefit from adaptive
selection. Under A2, adaptive achieves a mean Macro-F1 of 0.791,
compared with 0.833 for random, 0.809 for mean-bandwidth-matched
static, and 0.823 for fixed. For CNN1D-shape, adaptive reaches 0.702
versus 0.784 for static, a paired difference of $-0.0818$ with a
95\% confidence interval of $[-0.122,-0.041]$ and $p=0.0092$.
The controller also uses a more varied distribution of selected actions
on Sentinel than on CIC-IoT-2022, which is consistent with the greater
persistence of privacy after defense-aware training. This advantage is not uniform
across architectures, however: Transformer-shape is nearly unchanged
between adaptive and static camouflage (0.804 versus 0.801).

UNSW exhibits a different tradeoff. A 20\% packet-splitting
transformation is particularly effective for this traffic distribution,
allowing the mean-bandwidth-matched static defense to achieve lower
Macro-F1 than adaptive under both A0 (0.643 versus 0.723) and A2
(0.799 versus 0.834). For this comparison, the static duty schedule was selected from \textsc{Calibration} to match the adaptive controller's mean bandwidth
expenditure. The adaptive and static values reported here are taken
from the same paired comparison, accounting for the slight difference
between the adaptive UNSW mean here and the overall value reported in
RQ1. On \textsc{Test}, average bandwidth overhead remains similar at 5.18\% for static and 4.88\% for adaptive. However, 25.76\% of static \textsc{Test} windows
exceed the Balanced per-window bandwidth limit, compared with only
0.088\% under adaptive selection. Thus, the static defense provides
stronger fingerprint suppression on UNSW, while the adaptive controller
maintains substantially tighter per-window resource compliance.

Taken together, these results indicate that the benefit of adaptive
selection depends on whether the traffic distribution favors different
actions across windows or is well served by one consistently effective
transformation. IoT Sentinel retains the clearest defense-aware benefit
from adaptive selection, CIC-IoT-2022 shows greater learnability after
defense exposure, and UNSW highlights the value of resource-aware
control when a strong static transformation exists.

\subsection{How Does the Adaptive Controller Behave?}
\label{sec:controller_behavior}

\noindent\textbf{RQ3: Does the controller adapt its obfuscation
decisions to traffic context rather than collapse to a fixed behavior?}

Table~\ref{tab:controller_behavior} summarizes the Balanced
controller's behavior on \textsc{Test}. The context-aware
recommendation passes the confidence guard on 39.4--47.4\% of windows
across the three datasets. In many of these cases it agrees with the
global selector; it changes the selected action on 15.3\% of
CIC-IoT-2022, 24.6\% of IoT Sentinel, and 40.2\% of UNSW windows.
This shows that the contextual path often reinforces the global
decision, while still providing meaningful overrides when the detailed
traffic context supports a different action.

\begin{table}[t]
\centering
\caption{Balanced controller behavior on \textsc{Test}.}
\label{tab:controller_behavior}
\footnotesize
\setlength{\tabcolsep}{2.8pt}
\renewcommand{\arraystretch}{1.10}

\begin{tabularx}{\columnwidth}{
@{}l
c
c
c
>{\raggedright\arraybackslash}X@{}}
\toprule
\textbf{Dataset} &
\textbf{Context} &
\textbf{Changed} &
\textbf{No trans.} &
\textbf{Dominant action} \\
\midrule

CIC-IoT-2022
& 39.4\%
& 15.3\%
& 8.2\%
& 10\% padding (75.0\%) \\

IoT Sentinel
& 47.4\%
& 24.6\%
& 3.9\%
& 10\% packet splitting (66.4\%) \\

UNSW
& 43.8\%
& 40.2\%
& 18.9\%
& 5\% padding + 5\% packet splitting (40.9\%) \\

\bottomrule
\end{tabularx}

\vspace{2pt}

\parbox{\columnwidth}{\scriptsize
\textit{Note:} Context is the fraction of windows for which the
confidence guard accepts the context-aware recommendation; Changed is
the fraction for which that recommendation alters the global
selector's action. No trans. denotes an explicit no-transformation
decision.}
\end{table}

The controller's action choices also vary across datasets.
CIC-IoT-2022 is dominated by 10\% padding, while IoT Sentinel
primarily selects 10\% packet splitting. UNSW exhibits a broader
distribution of selected actions and leaves 18.9\% of windows
unchanged. These distributions show that adaptation need not appear as
frequent switching among actions: a single action may remain dominant
over many windows, while the controller can still select alternatives
or no transformation when the observed traffic context warrants a
different decision.

\subsection{Defense-Aware Learnability and Attacker Recovery}
\label{sec:defense_aware}

\noindent\textbf{RQ4: How much protection remains once the attacker
learns the defended traffic distribution?}

\begin{figure}[t]
    \centering
    \includegraphics[width=\columnwidth]
    {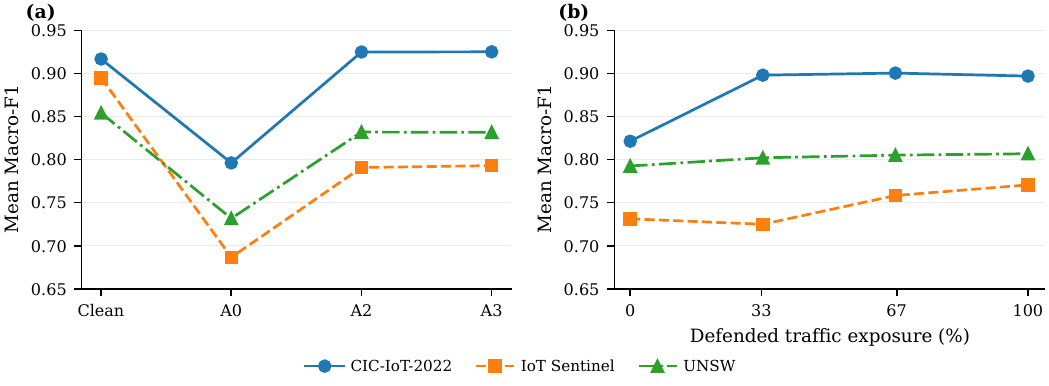}
    \caption{Defense-aware learnability and attacker recovery under the
    Balanced profile. (a) Mean Macro-F1 across the five core
    traffic-shape attackers on clean traffic and under adaptive
    camouflage in A0, A2, and A3. A2 trains a fresh attacker using
    defended adaptation traffic, whereas A3 evaluates the same A2
    attacker on a fresh stochastic realization of the adaptive
    defense without further training. (b) Mean Macro-F1 of the
    CNN1D-shape and Transformer-shape attackers as A4 progressively
    exposes them to defended adaptation traffic; 0\% denotes the
    clean-trained checkpoint. Lower Macro-F1 indicates stronger
    privacy.}
    \label{fig:defense_aware}
\end{figure}

Figure~\ref{fig:defense_aware}(a) shows that defense-aware training has
markedly different effects across datasets. Mean Macro-F1 rises from
0.796 to 0.925 on CIC-IoT-2022 and from 0.732 to 0.832 on UNSW,
recovering much of the performance lost under A0. IoT Sentinel retains
a larger privacy gap: Macro-F1 increases from 0.687 to 0.791 but
remains below the clean-traffic value of 0.895. A3 produces essentially
the same results as A2 on all three datasets (0.925, 0.793, and 0.832),
showing that a fresh stochastic realization of the same adaptive policy
does not restore protection once its defended distribution has been
learned.

\noindent\textbf{RQ5: How quickly does an attacker recover as defended
traffic becomes available?}

Figure~\ref{fig:defense_aware}(b) tracks CNN1D-shape and
Transformer-shape attackers as they are incrementally exposed to
defended traffic. Recovery is fastest on CIC-IoT-2022, where mean
Macro-F1 increases from 0.821 to 0.898 after roughly one-third of the
available exposure and changes little thereafter. IoT Sentinel shows
slower, non-monotonic recovery, reaching 0.771 at full exposure from
an initial value of 0.731. UNSW changes only modestly, from 0.793 to
0.807, with the sequence attackers already comparatively effective
before adaptation. These trajectories show that both the learnability
of defended traffic and the rate of attacker recovery vary across
traffic distributions.

\subsection{Cost of Causal Decision Making}
\label{sec:causality_gap}

\noindent\textbf{RQ6: How much privacy is lost because the controller
uses previous-window rather than complete current-window information?}

The deployed controller selects the action for window $W_t$ using
information available through $W_{t-1}$, allowing camouflage to be
applied as $W_t$ is transmitted. We compare this causal design with a
non-causal same-window reference that observes the complete current
window before selecting its action and would therefore require
full-window buffering.

Across RF, HGB, and MLP, access to complete current-window information
provides only a modest additional reduction in mean Macro-F1: 0.043 on
CIC-IoT-2022, 0.014 on IoT Sentinel, and 0.045 on UNSW. The two
controllers have slightly different realized communication costs, so
this comparison measures the benefit of more timely traffic information
rather than a cost-matched performance difference. Overall, the small
gap shows that previous-window control retains most of the aggregate
privacy benefit of complete current-window information without requiring
the traffic being protected to be buffered first.

\subsection{Protection Boundary under Richer Metadata}
\label{sec:metadata_boundary}

\noindent\textbf{RQ7: How much identifying information remains when
the attacker observes metadata beyond the traffic-shape features
targeted by camouflage?}

Across the five metadata-rich attackers, protocol-, service-, endpoint-,
flag-, and flow-level information remains available in addition to the
traffic characteristics modified by the controller. Under A0, Balanced
camouflage reduces mean Macro-F1 from 0.913 to 0.869 on CIC-IoT-2022,
from 0.911 to 0.861 on IoT Sentinel, and from 0.947 to 0.934 on UNSW.
These values remain substantially higher than the corresponding
traffic-shape results in Section~\ref{sec:privacy_cost}, showing that
metadata outside the transformed traffic-shape surface provides
additional identifying information.

After defense-aware training, mean Macro-F1 reaches 0.918, 0.898, and
0.958 on CIC-IoT-2022, IoT Sentinel, and UNSW, respectively. The
protection provided by the current controller is therefore specific to
the traffic-shape surface it modifies; protecting protocol-, service-,
and flow-level information would require complementary mechanisms.

\section{Discussion and Limitations}
\label{sec:discussion}

The results show that adaptive camouflage is most useful as a
resource-aware mechanism for choosing protection according to observed
traffic leakage. Its benefit after defense exposure varies across
datasets. IoT Sentinel retains the clearest advantage from adaptive
selection, whereas CIC-IoT-2022 and UNSW become substantially more
learnable once attackers are trained on defended traffic. Fresh
stochastic realizations provide little additional protection after the
defended distribution has been learned, and incremental exposure shows
that attacker recovery also varies across datasets. These findings indicate that action diversity alone does not ensure persistent protection; persistence also depends on how readily the defended traffic distribution can be learned.
The UNSW comparison further shows that a strong static action can
provide greater privacy at similar mean bandwidth, although with much
weaker per-window budget compliance.

The causal design retains most of the privacy benefit observed with the
same-window reference while avoiding full-window buffering. Because
the action for window $W_t$ is selected from information available
through $W_{t-1}$, changes observed during $W_t$ affect the subsequent
decision rather than the action already in progress. The protection
scope is also limited to the traffic characteristics manipulated by the
controller. Packet size, timing, direction-related behavior, and
packetization are modified, while protocol-, service-, endpoint-, and
flow-level information is not systematically concealed. The
metadata-rich experiments therefore identify a clear boundary of the
current design and motivate complementary protection when these signals
are available to the observer.

The evaluation uses public traffic traces, which enables controlled and
repeatable comparison across defenses, but does not capture every
effect that may arise in a live deployment. Resource feasibility is
estimated from conservative calibration-derived cost envelopes rather
than formal worst-case bounds, so occasional per-window deviations can
occur. The controller selects from a predefined set of camouflage actions,
so its achievable privacy--cost tradeoff also depends on the
effectiveness of those actions for a given traffic distribution.
More broadly, the defense-aware results motivate future policies that
vary the defended traffic distribution over time while continuing to
satisfy explicit bandwidth and latency constraints.

\section{Conclusion}
\label{sec:conclusion}

This paper presented Adaptive Traffic Camouflage, a causal,
leakage-aware, and resource-aware controller for reducing IoT device
fingerprinting on the traffic-shape surface. The controller selects
budget-feasible camouflage actions using previously observed traffic,
allowing protection to adapt to both leakage and deployment constraints.
Across three IoT datasets, the Balanced profile substantially reduces
fingerprintability at modest communication cost, while the Privacy
profile shows that larger resource budgets can enable considerably
stronger protection. Defense-aware evaluation further shows that the
persistence of this protection depends on the traffic distribution,
while the same-window comparison indicates that causal
previous-window control retains most of the privacy benefit observed
with complete current-window information without requiring full-window
buffering. Overall, the results show that adaptive camouflage can
allocate traffic-shape protection according to available communication
resources while preserving causal operation.

\bibliography{bibtext}

\bibliographystyle{IEEEtran}

\end{document}